\documentclass[%
 reprint,
superscriptaddress,
 amsmath,amssymb,
 aps,
]{revtex4-2}

\usepackage{color}
\usepackage[colorlinks=true, allcolors=blue]{hyperref}
\usepackage{graphicx}
\usepackage{dcolumn}
\usepackage{bm}
\usepackage{hyperref}

\begin{document}

 \title{Dislocation structure and mobility in the layered semiconductor InSe: \\ A first-principles study} 

\author{A.~N. Rudenko}
\email{a.rudenko@science.ru.nl}
\author{M.~I. Katsnelson}
\affiliation{Radboud University, Institute for Molecules and Materials, NL-6525 AJ Nijmegen, The Netherlands}
\affiliation{\mbox{Department of Theoretical Physics and Applied Mathematics, Ural Federal University, 620002 Ekaterinburg, Russia}}

\author{Yu.~N. Gornostyrev}
\affiliation{\mbox{M.N. Mikheev Institute of Metal Physics UB RAS, 620137, S. Kovalevskaya str. 18, Ekaterinburg, Russia}}
\affiliation{\mbox{Department of Theoretical Physics and Applied Mathematics, Ural Federal University, 620002 Ekaterinburg, Russia}}

\date{\today}

\begin{abstract}
The structure and mobility of dislocations in the layered semiconductor InSe is studied within a multiscale approach based on generalized 
Peierls--Nabarro model with material-specific parametrization derived from first principles. 
The plasticity of InSe turns out to be attributed to peculiarities of
the generalized stacking fault relief  for the interlayer dislocation slips
such as existence of the stacking fault with a very low energy and low
energy barriers.
Our results give a consistent microscopic explanation of recently observed [Science {\bf 369}, 542 (2020)] exceptional plasticity of InSe.
\end{abstract}


\maketitle

\section{Introduction}

Two-dimensional (2D) semiconductors have attracted the great interest due to their  electrical and optical properties that make them promising candidates for future electronic and opto-electronic applications \cite{Wang2012,Mak2016,Sanchez2014}.

Indium selenide (InSe) is a typical representative of the layered van der Waals (vdW) semiconductors, which has received considerable attention in recent years because of the possibility to fabricate InSe structures down to monolayer thickness. Atomic-thick InSe crystals demonstrate high environmental stability and attractive electronic properties \cite{Chen2014,Mudd2016,Bandurin2017,Patane2017,Lugovskoi2019}, appealing for practical applications. 

The presence of a vdW gap between different layers makes the material easy to cleave, thus making it difficult to retain its structural integrity through plastic deformations.
As a consequence, plastic deformability is highly unexpected for bulk vdW inorganic semiconductors at room temperature. However, recent studies demonstrate a number of counterexamples \cite{Han2007,Shi2018,Wei2020,Zhao2019}.
In fact, the ability to sustain large deformations without fracture is an exceptional property of layered semiconductors \cite{Wei2020,Zhao2019}, which makes them advantageous for applications 
compared to conventional organic semiconductors.
Apart from vdW inorganic semiconductors, high plasticity is typical to layered carbides and nitrides, constituting a broad class of materials, the so-called MAX-phases \cite{MAXphase1,MAXphase2}.

Mechanical properties of InSe are characterized by superior in-plane flexibility of individual layers \cite{Zhao2019}. 
As it has been recently shown \cite{Wei2020}, layered phase of InSe demonstrates exceptional plasticity among other vdW inorganic semiconductors. 
This behavior can be attributed to the low resistance of InSe with respect to the interlayer gliding.
The origin of such unusual behavior has been discussed in Ref.~\onlinecite{Wei2020} in terms of the cohesion characteristics of vdW inorganic semiconductors, and anomalously low Young's modulus of individual layers \cite{Zhao2019}. 
As suggested in Ref.~\onlinecite{Wei2020}, soft intralayer bonding between In and Se atoms may be responsible for rather low energy barrier to slipping between the \{001\} layers. At the same time, the microscopic mechanism of such an exceptional behavior is not yet understood, and a quantitative theory of the high plasticity observed in InSe is still missing.

It should be noted that a similar mechanism of plasticity is realized in MAX-phases \cite{MAXphase1}, 
in which slipping of close-packed planes is governed by the motion of dislocations in the basal plane \{001\},
facilitated by weak interlayer bonding.
Although MAX-phases are characterized by the localization of deformations along with the formation of kink bands, 
the dislocation motion plays a key role in this process.
At the same time, the criteria determining the regime of plasticity typical to true bulk materials remain applicable even for layered MAX-phases \cite{Thompson2018}. 

In this study, we employ a multiscale theoretical approach to analyze
the structure and mobility of dislocations as well as brittle vs. ductile behavior of InSe. 
We start from an \emph{ab initio} description of the 
interatomic interactions 
to retrieve the generalized 
stacking fault (GSF) energy surface, 
which is further used to calculate the dislocation properties 
within the generalized Peierls–Nabarro (PN) model. 

We show that the GSF energy surface of InSe is characterized by a very low energy barrier and existence of the stacking fault with almost zero energy. This leads to a large splitting of the dislocation core and exponential suppression of the Peierls stress. This is sufficient to provide a ductile behavior of a single-crystalline layered InSe.

\section{Theoretical approach and computational details}

\begin{figure*}[tb]
	\begin{center}
		\includegraphics[width=13.50cm]{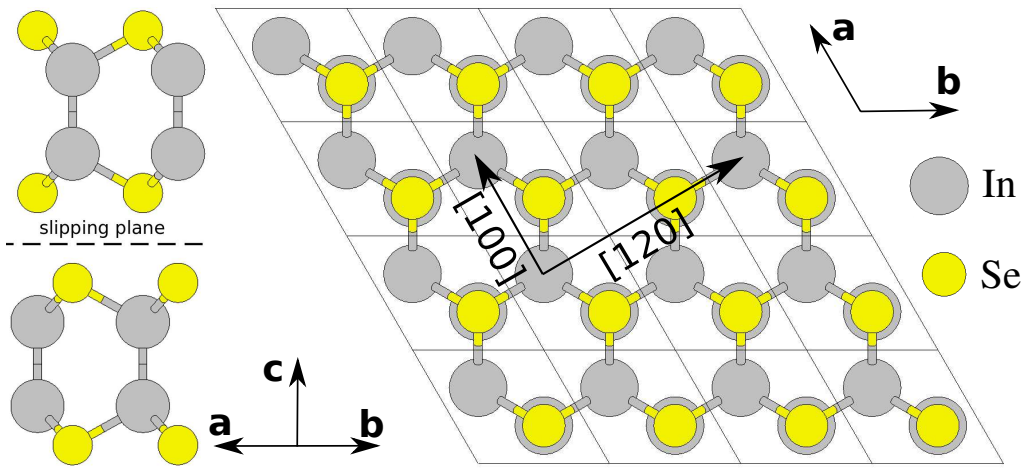} \ \ \ \ \ \ \ \
		\caption{Schematic representation of the InSe crystal structure (AA-stacking) shown in two projections. The arrows in the (001) plane denote [100] (zigzag) and [120] (armchair) nonequivalent directions in which the slipping of one layer is considered relative to the other.
		}
		\label{fig1}
	\end{center}
\end{figure*}

Peierls-Nabarro (PN) model provides a basis of our understanding of dislocation properties which determines strength and plasticity of materials \cite{Hirth}. A combination of \emph{ab initio} calculations and the PN model \cite{Mryasov1996,Mryasov1998} 
allows one to describe the 
dislocation structure starting from accurate electronic structure calculations of specific materials. The key point is to calculate the energetics of GSF from the first principles. 
This approach has been successfully used to analyze of the structure and mobility of dislocations in metals and alloys
in a large number of realistic situations 
\cite{Mryasov1998,yugIr,Mryasov2001,Mryasov2002,Wang2011,Liu2017,Kamimura2018,Xu2020}. 
In the framework of the generalized PN model, the total energy of the crystal with an infinite straight dislocation  can be written as \cite{Hirth}
\begin{equation}
E_{tot}[{\bf u}(x)] = E_{el}[{\bf u}(x)] + E_{mis}[{\bf u}(x)],
\label{eq:toten}
\end{equation}
where $E_{el}$ is the elastic energy term, and $E_{mis}$ is the nonlinear atomistic misfit energy term, both depending on relative displacements
${\bf u}(x)$ of atom rows below and above the cut plane, with $x$ being the distance from the dislocation axis in the slip plane 
The misfit energy
\begin{equation}
E_{mis} = h \sum_n \Phi [{\bf u}(nh - l)]
\label{eq:misen}
\end{equation}
is defined by a periodic energy profile $\Phi({\bf u})$ where $n$ is the number of atomic rows, $h$ is the distance between them, and $l$ is the position of the dislocation center in the lattice. The function $\Phi({\bf u})$ is usually approximated by the GSF 
energy or the $\gamma$-surface \cite{Vitek}, which is associated with shearing half of the crystal by a vector ${\bf u}$ in the slip plane.
Within the PN model, the displacement field ${\bf u}(x)$ is determined from the balance condition for the elastic stress and
the atomistic forces originating from the interaction between two half-spaces of the crystal (see Ref.~\onlinecite{Hirth} for details).
Within the PN model, this balance condition is determined by
the minimum of the functional 
Eq.~(\ref{eq:toten}), $\delta E_{tot}/\delta {\bf u}(x)=0$.

In order to estimate the resistance of InSe with respect to shear deformation, we follow the GSF scheme and
construct a series of different stacking configurations of bilayer InSe corresponding to a shearing of one layer relative 
to the other. We then calculate the energy profiles $\Phi [{\bf u}(x)]$ along two sliding pathways: [100] (zigzag) and [120] (armchair), 
as it is shown in Fig.~\ref{fig1}. The energies of various structural 
configurations were calculated using density functional theory. The calculations were carried out using the 
projected augmented wave formalism \cite{Blochl94} as implemented in the \emph{Vienna ab initio simulation package} ({\sc vasp}) 
\cite{Kresse1,Kresse2}. The exchange-correlation effects were taken into account by using the dispersion-corrected nonlocal 
functionals, vdW-DF \cite{Dion2004} and opt-B88-vdW \cite{Klimes2011}. For comparative purposes, we also used the generalized gradient approximation (GGA) \cite{pbe} and Hartree-Fock approximation (HF) in the form of the exact exchange energy \cite{HF} to calculate the interlayer energies. We used hard pseudopotentials, which include 4$s$, 4$p$, 4$d$, 5$s$, 5$p$ valence 
electrons for In, and 2$s$, 2$p$, 3$d$, 4$s$, 4$p$ electrons for Se. A 900 eV energy cutoff for the plane-waves and a convergence threshold 
of 10$^{-8}$ eV were used. The Brillouin zone was sampled by a (16$\times$16) {\bf k}-point mesh. In order to avoid interactions between the supercell images in the nonperiodic direction, a 30 \AA~ thick vacuum slab was added in the direction normal to the InSe sheet. 
In the calculations of slipping energies, the atomic positions were relaxed in the vertical direction until the residual forces were 
less than 10$^{-3}$ eV/\AA. In the calculations with variable interlayer separation, one of the atoms in each layer was fixed. The 
lateral lattice constant was fixed to the equilibrium value $a = 4.0$ \AA. 

\section{Results}

\subsection{Interlayer interaction}

In this study, we consider InSe crystal, which is composed of vertically stacked weakly interacting InSe layers. Each layer
adopts a crystal structure with two vertically displaced
2D buckled honeycombs (Fig.~\ref{fig1}). Symmetry of each layer is characterized by the point group $D_{3h}$. In accordance with Ref.~\onlinecite{Wei2020}, we assume that InSe crystallizes in the $\beta$-InSe phase with the space group P6$_3$/mmc (No. 193), in which In and Se atoms in the adjacent layers are vertically aligned. 

\begin{figure}[tbp]
	\begin{center}
		\includegraphics[width=9.0cm]{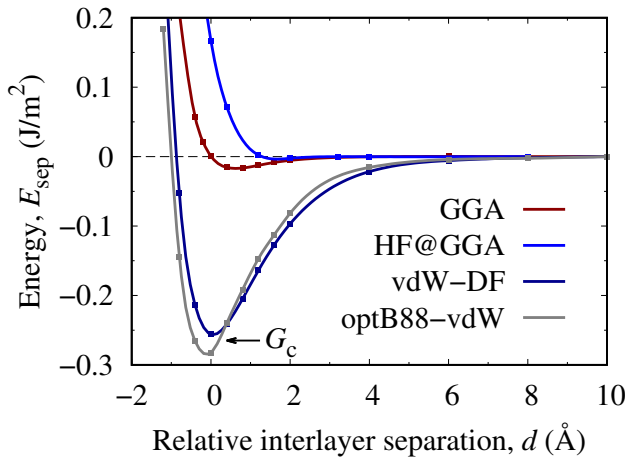} 
		\caption{The energy of bilayer InSe calculated as a function of the interlayer separation $d$ using different exchange-correlation functionals. $d=0$ corresponds to the equilibrium distance. $G_c=-E_{\mathrm{sep}}(0)$ denotes the cleavage energy.}
		\label{fig5}
	\end{center}
\end{figure}

The calculated separation energy $E_{\mathrm{sep}}$ as a function of the distance between the two crystal layers cleaved along the plane
\{001\} is presented in Fig.~\ref{fig5}. The results in Fig.~\ref{fig5} are shown for two typical dispersion-corrected exchange-correlation functionals, vdW-DF and optB88-vdW, in comparison with the standard generalized-gradient approximation (GGA) and Hartree-Fock (HF) calculations. Both GGA and HF binding energy curves demonstrate a shallow minimum, indicating the absence of any sizeable binding between the layers. This demonstrates that the interlayer binding in InSe is mainly of the vdW origin. On the contrary, nonlocal vdW-functionals predict a considerable interlayer interaction, typical to layered materials. The resulting cleavage energy $G_c=-E_{sep}(0)$
amounts to 0.26 and 0.29 J/m$^2$ for vdW-DF and optB88-vdW functionals, respectively. Both functionals result in similar $E_{\mathrm{sep}}(d)$ 
behaviour. In the discussion that follows, we consider the results obtained using the vdW-DF functional only. 

\subsection{Generalized stacking fault energy}

\begin{figure*}[tbp]
	\begin{center}
		\includegraphics[width=7.840cm]{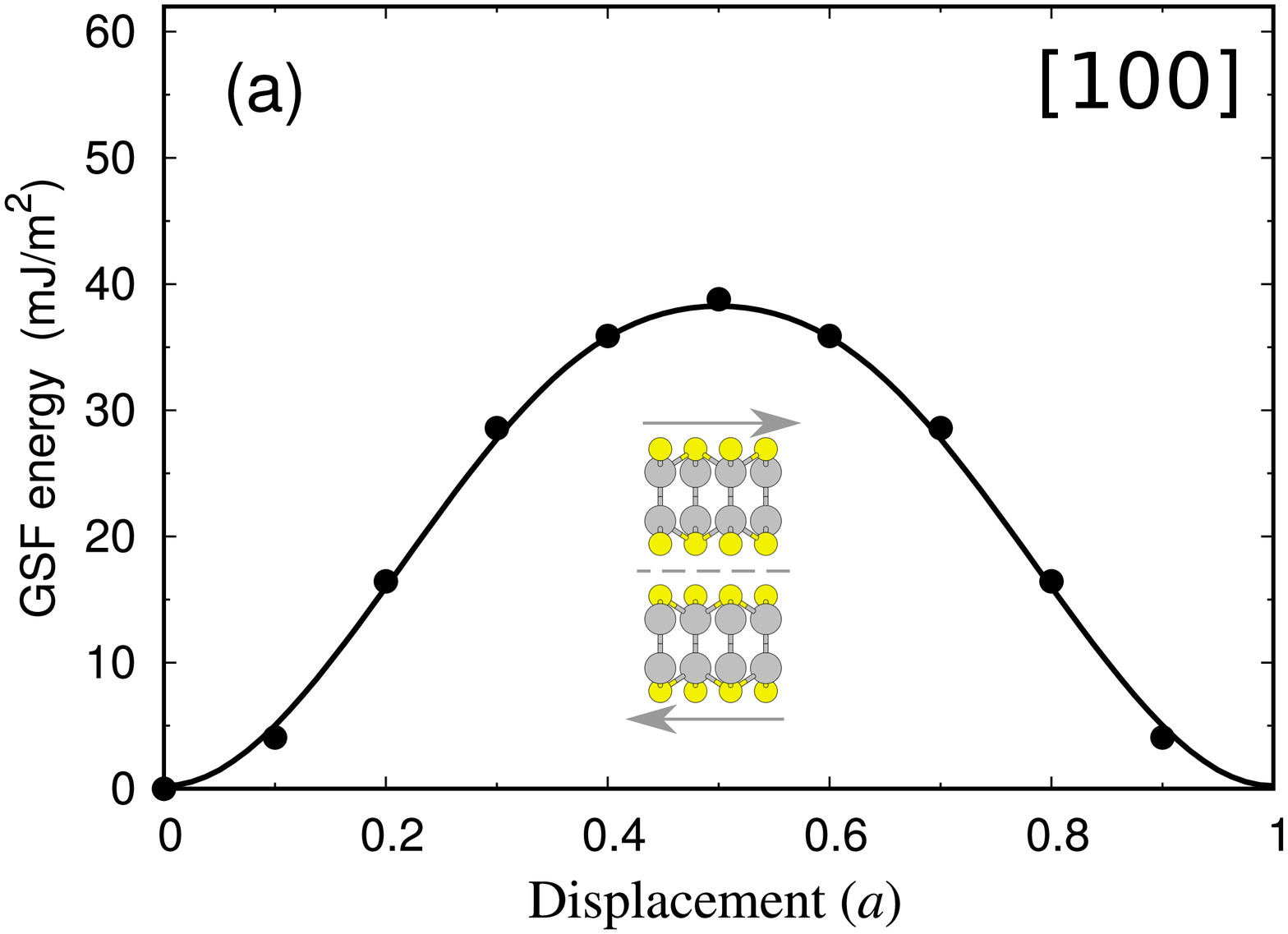} \ \ \ 
		\includegraphics[width=7.840cm]{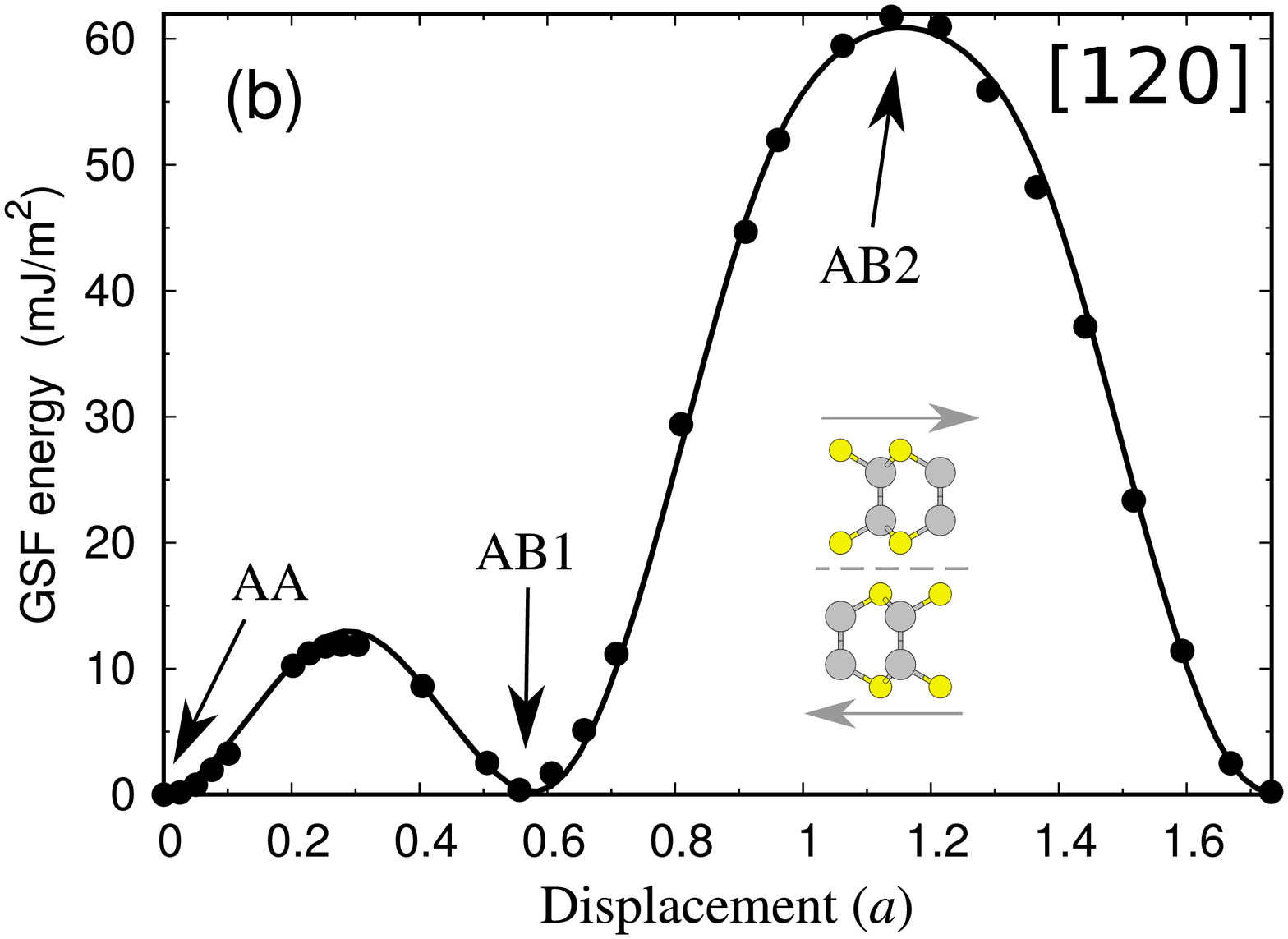}
		\caption{Cross-sections of the GSF energy along (a) [100] (zigzag) and (b) [120] (armchair) directions in the (001) plane.
		Points correspond to the results of \emph{ab initio} calculations within the nonlocal vdW-DF functional. Solid lines show the approximation obtained by fitting with Eq.~(\ref{gsf:apr}). Arrows denote symmetric configurations corresponding to AA- and AB-types of stacking of the two layers.}
		\label{fig2}
	\end{center}
\end{figure*}

The energy profiles $\Phi ({\bf u})$ calculated along two sliding pathways, corresponding to the [120] (armchair) and [010] (zigzag) cross-sections of the GSF surface, are presented in Fig.~\ref{fig2}. One can see that the sliding along [120] between the AA and AB1 configurations is the most energetically favorable. The corresponding energy barrier is around $\sim$14 mJ/m$^2$.
Interestingly, the energies of the initial (AA) and final (AB1) configurations are very close to each other.
This means that the AA--AB1 sliding results in the formation of a stacking fault with nearly zero energy.
This is due to the fact that the main contribution to the energy upon sliding of InSe layers is given by the adjacent Se atoms, which have equivalent positions in both AA and AB1 stacking configurations. 
As a consequence, the most plausible mechanism of plasticity in InSe is the motion of partial dislocations with the Burgers vector ${\bf b}_{par}$=AA--AB1 in the \{001\} plane.

It is convenient to perform a Fourier expansion of the GSF energy over the reciprocal vectors of a 2D lattice in the \{001\} plane, which can be written as
\begin{eqnarray}
\label{gsf:apr}
\Phi({\bf u}) = A_0 + \sum_{k=1}^{k_{max}}  \sum_{i=1}^3 A_k \cos(k{\bf b}_i{\bf u})+ \quad \quad \quad \quad \quad  \\
\notag
\sum_{k=1}^{k_{max}}  \sum_{i=1}^3 A_k \cos(k{\bf b}_i({\bf u}+{\bf u}_0))
\end{eqnarray}
where $A_k$ are the coefficients determining the entire GSF surface.
Here ${\bf b}_1= 2 \pi /3 (1,\sqrt{3},0)$, ${\bf b}_2= 2 \pi /3 (1,-\sqrt{3},0)$ are the basis vectors of the reciprocal lattice in units of the inverse lattice constant $a^{-1}$, and ${\bf b}_3= {\bf b}_1+{\bf b}_2$.
The second and third terms in Eq.~(\ref{gsf:apr}) describe contributions from two triangular lattices shifted by the vector ${\bf u}_0=(1,0,0)$ relative to each other, which ensures a hexagonal symmetry of the function $\Phi({\bf u})$ in the \{001\} plane.
The coefficients $A_k$ have been determined by fitting to the results of \emph{ab initio} calculations presented in Fig.~\ref{fig2}. The expansion length has been set to $k_{max}=3$, which is sufficient to obtain a reliable approximation of the calculated points.

\begin{figure}[t]
	\begin{center}
		\includegraphics[width=8.30cm]{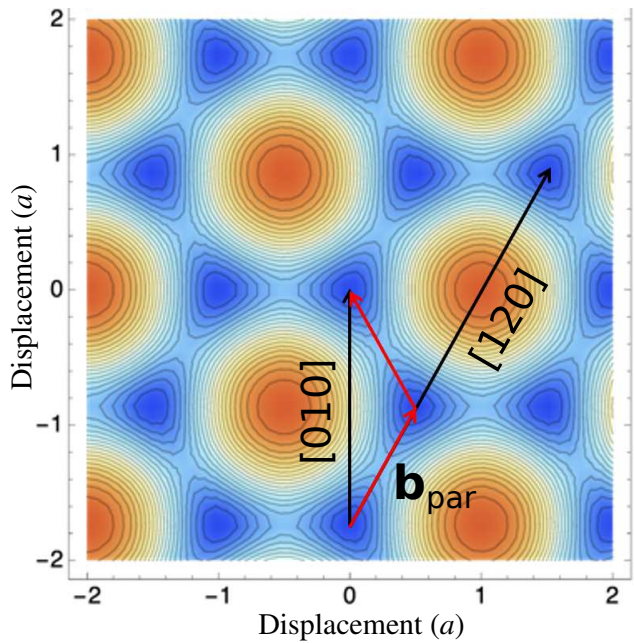} 
		\caption{The GSF surface obtained from the Fourier expansion Eq.~(\ref{gsf:apr}) with parameters fitted to the 
		calculated GSF cross-sections, presented in Fig.~\ref{fig2}. The black arrows correspond to [010] (zigzag) and [120] (armchair) directions. The red lines correspond to the Burgers vector of partial dislocations discussed in the text.		}
		\label{fig3}
	\end{center}
\end{figure}

The resulting approximation to the entire GFS surface is presented in Fig.~\ref{fig3}, with the individual cross-sections along the [010] (zigzag) and [120] (armchair) directions shown by the solid lines in Fig.~\ref{fig2}.
From Fig.~\ref{fig3} one can see that a direct translation by vector 
${\bf c}_2=(0,\sqrt{3},0)$ along [010], which was considered in Ref.~\onlinecite{Wei2020}, is not the most energetically favorable slipping path. 
Instead, the minimum energy path corresponds to a decomposition of  {\bf c}$_2$ into two partial translations as $(0,\sqrt{3},0) \rightarrow (1/2,\sqrt{3}/2,0)+(-1/2,\sqrt{3}/2,0)$.  
Since the partial translation $(1/2,\sqrt{3}/2,0)$ leads to the formation of a stacking fault with nearly zero energy (see Fig.~\ref{fig2}), we expect that the motion of partial dislocations with the Burgers vector ${\bf b}_{par}=(1/2,\sqrt{3}/2,0)$ in the \{001\} plane is the main mechanism of plastic deformations in InSe.

\subsection{Dislocation structure and mobility}

We now analyze the structure and mobility of partial dislocations with the Burgers vectors 
${\bf b}_{par}=(1/2,\sqrt{3}/2,0)$ 
in the \{001\} plane.
To determine the dislocation core structure, we employ a modified PN model as it is discussed in Refs.~\onlinecite{Mryasov1996,Mryasov1998}. 
We perform a minimization of the total energy functional, Eq.~(\ref{eq:toten}), with a discrete representation of the misfit energy, 
Eq.~(\ref{eq:misen}), using trial functions, ${\bf u}(x)$, defined from the Laurent expansion  \cite{Leicek,Mryasov1998}
of their derivatives
\begin{equation}
\begin{aligned}
\rho(x) = \frac {du(x)}{dx} &= \frac {1}{2} \sum_{n=1}^{p} \left[ \frac {A_n}{(x-i\zeta)^n} +\frac {\bar{A}_n}{(x+i\zeta)^n} \right], \\
\frac{2}{\mu D}  \frac {dE_{mis}[u(x)]}{dx} &= -\frac {i}{2} \sum_{n=1}^{p} \left[ \frac {A_n}{(x-i\zeta)^n} -\frac {\bar{A}_n}{(x+i\zeta)^n} \right],
\label{gsf:PNe}
\end{aligned}
\end{equation}
where $\mu$ is the shear modulus, 
$D$ is a parameter equal to 1 for a screw dislocation, and to 1/(1-$\nu$) for an edge dislocation with $\nu$ being the Poisson ratio,
and $\zeta$ determines the width of the dislocation core. It is sufficient to use $p=2$ for the case under consideration. Using Eq.~(\ref{gsf:PNe}), the expressions for displacements ${\bf u}(x)$ and restoring forces ${\partial E_{mis}[u(x)]}/{\partial u}$ can be represented in the conventional form as
\begin{equation}
\begin{aligned}
\frac {2 \pi u(\theta)}{b} & = \theta - \frac {\alpha -1}{\alpha} \sin\theta +\beta \sin\theta, \\
\frac{4 \pi \omega_0}{\mu Db}  \frac {dE_{mis}}{du} &= \frac {1}{\alpha} \left[ \sin\theta + \frac {\alpha -1}{\alpha}
2\sin^2\frac {\theta}{2} \sin\theta + \right. \\ & \left. 2\beta\sin^3\frac{\theta}{2}\cos\frac {3\theta}{2} \right],
\label{gsf:PNe2}
\end{aligned}
\end{equation}
where $\theta=2\,\mathrm{arctan} (x/\zeta)$, $\zeta = \alpha \omega_0$, $\omega_0=dD/2$ is the width of the dislocation core
defined within the original PN model \cite{Hirth}, and $d$ is interplane distance.

\begin{figure}[t!b]
	\begin{center}
		\includegraphics[width=8.70cm]{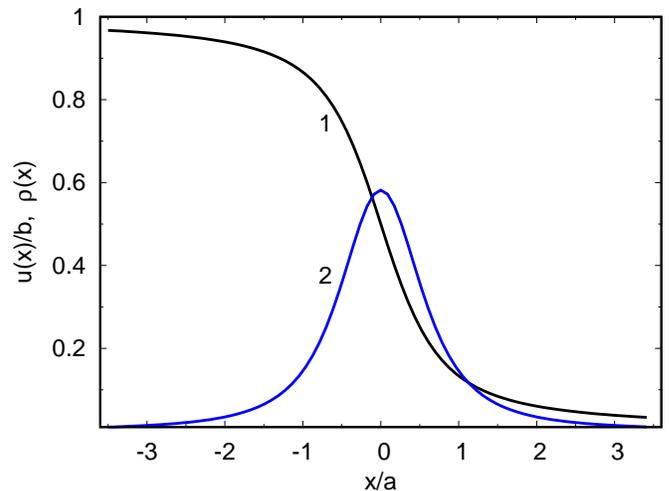} 
		\caption{ Structure of the core of screw dislocation with 
		{\bf b}$_{par}$ Burgers vector.  Distribution of displacements $u(x)/b$ (curve 1) and 
		density of infinitesimal dislocations $\rho (x)$ (curve 2).}
		\label{fig4}
	\end{center}
\end{figure}

The parameters $\alpha$ and $\beta$ have been determined from fitting Eq.~(\ref{gsf:PNe2}) to the calculated values of $E_{mis}(u)$ in the interval AA--AB1 (Fig.~\ref{fig2}). 
The resulting distribution of displacements $u(x)$ and $\rho(x)$ for
screw dislocation are shown in Fig.~\ref{fig4}.
One can see that the dislocation under consideration has a rather wide core, facilitating gliding processes.

Let us estimate the Peierls stress, $\sigma_P$, which determines the lattice resistance to the dislocation motion.
Within the generalized PN model considered in this work, $\sigma_P$ can be represented as \cite{Mryasov1996}
\begin{equation}
\sigma_P = \frac {2S_0}{bh} \int_{-\infty}^{\infty} \frac {\partial E_{mis}}{\partial u} \, \frac {\partial u}{\partial x} \,
\sin\left(2\pi x\right)dx,
\label{gsf:PNs}
\end{equation}
where $S_0$ is the area per atom in the slip plane. The integral Eq.~(\ref{gsf:PNs}) can be approximated using the method of steepest descent, yielding
\begin{equation}
\sigma_P \approx \frac {S_0}{h\sqrt{\pi}} \exp\left(-\frac {\pi \alpha \omega_0}{h}\right)\left(\frac {\partial E_{mis}}{\partial u}\right)\bigg\rvert_{x=0}.
\label{gsf:PNs2}
\end{equation}
Using the calculated GSF energy, we obtain $\sigma_P /\mu \approx$ 4.7$\times$10$^{-5}$.
Therefore, the dislocations under consideration are characterized by considerably low Peierls stress and, as a consequence, high mobility typical to ductile metals (see Ref. \onlinecite{yugIr}).

\subsection{Cleavage fracture condition}

Rice and Thomson \cite{RT74} suggested an approach to characterise whether a material is ductile or brittle by considering the balance of the crack opening and dislocation emission from the crack tip, processes that characterize the microscopic properties of materials.
The emission of a dislocation can cause crack blunting, lowering the strain energy release rate and thus leading to ductile behavior.
According to the Rice-Thompson criteria, brittle failure is realized 
when $\mu b/\gamma >7.5-10$, where $\gamma =G_c/2$ is the 
surface energy. The applicability of this criterion for the ranking of materials with different crystal structures has been widely discussed 
(see Ref.~\onlinecite{Thompson2018}).

From the results presented in Fig.~\ref{fig5}, we have $\gamma=0.13$ J/m$^2$, which is considerably smaller compared to typical hexagonal close-packed metals with $\gamma =$ 1-2 J/m$^2$ \cite{Luo2014}. 
However, because of the low shear modulus in the \{001\} plane $\mu=1.75$ GPa, we obtain $\mu b/\gamma \approx 3.2$. Such a small ratio $\mu b/\gamma $ indicates that 
brittle fracture along the \{001\} plane is highly unlikely.

Another important parameter of fracture is the ratio $\gamma /\gamma_{us}$ \cite{R92}, where $\gamma_{us}$ is the so-called 
unstable stacking fault energy, which characterizes a barrier for sliding. In the case under consideration $\gamma /\gamma_{us}$ has its maximum along the AA -- AB1 path
in Fig.~\ref{fig2}(b). An estimation gives rather high ratio $\gamma /\gamma_{us} \approx$ 8.3, which is similar to the values typical for ductile face-centered cubic metals Cu and Ni \cite{R92}.
It should be noted that the plasticity of layered crystals was discussed in Ref.~\onlinecite{Wei2020} in terms of the relation of slipping and cleavage energies, which is similar to the $\gamma /\gamma_{us}$ ratio.

Strictly speaking, the criteria discussed above should be applied with caution to layered crystals such as InSe or MAX phases (see discussion 
in Ref.~\onlinecite{Thompson2018}). For a correct description of plasticity in such materials,
the dislocation mobility must be taken into account. Since we have found that the intrinsic dislocation mobility in InSe is rather high, one can expect that the Rice-Thomson criterion correctly predicts the ductile behavior in this compound.\\

\section{Discussion and conclusion}

To elucidate the origin of ductile behavior of the 2D layered compound InSe an {\it ab initio} based analysis of the dislocation structure, mobility, and
cleavage decohesion was performed. We found that the AA–AB1 sliding in the $\langle120\rangle$ (armchair) direction is characterized by a very low energy barrier and results in the formation of stacking faults with nearly zero energy (see Fig. \ref{fig2}). 
As a result, partial dislocations with the Burgers vector ${\bf b}_{par}$ along 
$\langle120\rangle$ can easily 
glide in the basis plane.  We note that a similar behavior of the GSF energy surface was reported earlier for graphite \cite{Savini2011} 
where high mobility of dislocations in the basis plane is also to be expected \cite{Telling2003}.

Low energy barrier for the AA–AB1 sliding results in rather small stacking fault energy $\gamma_{us}$, resulting in the large ratio $\gamma/\gamma_{us}$, indicating a ductile character of InSe. This allows us to conclude about the intrinsic plasticity of InSe in the situation when the deformation is applied 
along the basal plane;
a similar conclusion can be drawn considering the ratio 
$\mu b/\gamma $ that appears in the Rice-Thomson criterion of the fracture condition.
Thus, monocrystalline InSe is microscopically ductile material. This behavior does not appear surprising for layered materials provided there is a preferred gliding direction (see discussion in Ref.~\onlinecite{Thompson2018}).
At the same time, this does not guarantee macroscopic plasticity of polycrystalline samples, since only one slipping plane cannot provide an arbitrary shape change of material. According to the von Mises criterion, five independent slip systems are required to provide an arbitrary deformation \cite{Hirth} which is highly improbable for layered materials with geometrically determined special directions. Therefore our analysis is sufficient only to explain ductility of single-crystalline materials under special kind of deformation, exactly what was observed experimentally \cite{Wei2020}.


 \end{document}